\documentclass[
 reprint,
 amsmath,
 pre,
 amssymb,
 nofootinbib,
 aps,
]{revtex4-1}

\usepackage[pdftex]{graphicx}
\usepackage{wasysym}
\usepackage{amsfonts}
\usepackage{ifsym}

\makeatletter
\newlength{\apb@width}
\newcommand{\autoparbox}[2][c]{\settowidth{\apb@width}{#2}\parbox[#1]{\apb@width}{#2}}

\makeatother

\newcommand{\namedref}[2]{\hyperref[#2]{#1~\ref*{#2}}}

\renewcommand{\Re}{\mathop{\mathrm{Re}}}

\newcommand{\Csphere}{{}^\bullet\kern-1.2pt C}
\newcommand{\Ctorus}{{}^\circ\kern-1.2pt C}

\newcommand{\COMMENT}[1]{}

\newcommand{\neqa}{\nonumber\end{eqnarray}}

\newcommand{\<}{{\langle}}
\renewcommand{\>}{{\rangle}}

\newcommand{\re}{\relax{\rm I\kern-.18em R}}

\def\su2{{SU(2)}}

\def\[{\left[}
\def\]{\right]}

\def\({\left(}
\def\){\right)}
\def\[{\left[}
\def\]{\right]}

\def\<{\langle}
\def\>{\rangle}

\def\i2{\frac{i}{2}}

\def\2F1{\,_2{\rm F}_1}

\usepackage[usenames,dvipsnames]{xcolor}
\usepackage{color}
\usepackage{graphicx}
\usepackage{dcolumn}
\usepackage{bm}
\usepackage{hyperref}

\newcommand{\beq}{\begin{equation}}
\newcommand{\eeq}{\end{equation}}
\newcommand{\beqq}{\begin{equation*}}
\newcommand{\eeqq}{\end{equation*}}
\newcommand\beqa{\begin{eqnarray}}
\newcommand\eeqa{\end{eqnarray}}
\newcommand\beqaa{\begin{eqnarray*}}
\newcommand\eeqaa{\end{eqnarray*}}
\newcommand\bea{\begin{array}}
\newcommand\eea{\end{array}}

\begin{document}


\title{Bootstrapping QCD:\\ the  Lake, the Peninsula and the Kink}

\author{Andrea L. Guerrieri}
\affiliation{Instituto de F\'isica Te\'orica, UNESP, ICTP South American Institute for Fundamental Research, Rua Dr Bento Teobaldo Ferraz 271, 01140-070, S\~ao Paulo, Brazil}
\author{Jo\~ao Penedones}
\affiliation{Institute of Physics, \'Ecole Polytechnique F\'ed\'erale de Lausanne (EPFL), \\
Rte de la Sorge, BSP 728, CH-1015 Lausanne, Switzerland}
\author{Pedro Vieira}
\affiliation{Instituto de F\'isica Te\'orica, UNESP, ICTP South American Institute for Fundamental Research, Rua Dr Bento Teobaldo Ferraz 271, 01140-070, S\~ao Paulo, Brazil}
\affiliation{Perimeter Institute for Theoretical Physics, 31 Caroline St N Waterloo, Ontario N2L 2Y5, Canada}


\begin{abstract}
We consider the S-matrix bootstrap of four dimensional scattering amplitudes with $O(3)$ symmetry and no bound-states. We explore the allowed space of scattering lengths which parametrize the interaction strength at threshold of the various scattering channels. Next we consider an application of this formalism to pion physics. A signature of pions is that they are derivatively coupled leading to (chiral) zeros in their scattering amplitudes. In this work we explore the multi-dimensional space of chiral zeros positions, scattering length values and resonance mass values. Interestingly, we encounter 
lakes, peninsulas and kinks 
depending on which sections of this intricate multi-dimensional space we consider. We discuss the remarkable location 
  where QCD seems to lie in these plots, based on various experimental and theoretical expectations.  \\ 

\centering \textit{In memory of Clay Riddell.}
\end{abstract}

\pacs{Valid PACS appear here}
\maketitle



\section{\label{sec:Setup}Introduction and Setup}

Pions are approximate Goldstone bosons for spontaneous chiral symmetry breaking in QCD. 
In this Letter we study the question: \emph{Do pion scattering amplitudes take a special place in the space of consistent S-matrices?}

We assume that the up and down quarks are massive and degenerate ($N_f=2$), a very good approximation of the physical world.~\footnote{Chiral symmetry is broken not only spontaneously $SU(2)_L \times SU(2)_R \to SU(2)_V$, but also explicitly by a mass term for the quarks. We neglect the $SU(2)_V$ violations coming from the $u{-}d$ quark mass difference and the electroweak interactions hence pions are stable particles in our setup. Finally, when plugging numbers in resonance masses, scattering lengths, etc we will use real world experimental data.}
To wit we consider pions as particles of mass $m_\pi=1$ in the vector representation of~$O(3)$ so that their 2 to 2 scattering amplitude reads 
\begin{eqnarray}
&&\mathcal{T}_{ab}^{cd} = 
A(s|t,u)\delta_{ab} \delta^{cd}+
A(t|s,u)\delta_a^c \delta_b^d+
A(u|s,t)\delta_a^d \delta_b^c \,, \nonumber
\end{eqnarray}
where $s,t,u=4-s-t$ are the usual Mandelstam invariants.
Crossing symmetry  is simply $A(s|t,u)=A(s|u,t)$.
The partial wave expansion is given by 
\begin{eqnarray}
&&\mathcal{T}  =\Big(3 A(s|t,u)+A(t|s,u)+A(u|s,t) \Big) \mathbb{P}_{0}  + \\
&&+ \Big(A(t|s,u)-A(u|s,t) \Big) \mathbb{P}_{1}  + \Big(A(t|s,u)+A(u|s,t) \Big) \mathbb{P}_{2}   \nonumber\\ 
&&  =\frac{16\pi i \sqrt{s}}{\sqrt{s-4}}\sum_{I=0,1,2}  \mathbb{P}_I \!\sum_\ell  (2\ell+1) \Big(1- {\color{blue} S_\ell^{(I)}(s)}\Big)P_\ell\!\left(\frac{u-t}{u+t}\right) \,, \nonumber
\end{eqnarray}
where $P_\ell$ are Legendre polynomials and  $\mathbb{P}_{I}$ are the ($s$-channel) projectors onto the three possible isospin channels.
($I=0$ for singlet, $I=1$ for vector and $I=2$ for symmetric traceless tensor).
Due to the absence of bound states, we consider the following analytic and crossing symmetric ansatz~\cite{Paulos:2017fhb}:
\begin{eqnarray}
\!\!\!\!\!\!\!\! A(s|t,u)\!=\!\!\sum^\infty_{n\le m} \!\! a_{nm} \!\big(\rho_t^n \rho_u^m\!{+}\!\rho_t^m \rho_u^n\big)\!+\!\sum^\infty_{n,m} \! b_{nm} \!\big(\rho_t^n\!{+}\!\rho_u^n\big) \rho_s^m 
,\label{ansatz} 
\end{eqnarray}
where $\rho_z \equiv \frac{\sqrt{8/3}-\sqrt{4-z}}{\sqrt{8/3}+\sqrt{4-z}}$ is a conformal mapping of the $z$ complex plane minus the cut $z>4$ to the $\rho$ unit disk.~\footnote{Indeed, all our results are independent on the conformal mapping chosen to foliate the complex plane. Other works used conformal mappings to parametrize fixed partial wave amplitudes~\cite{GarciaMartin:2011cn,Caprini:2008fc}. }
In the partial wave decomposition, unitarity reads simply~$|S^{(I)}_\ell(s)|\le 1$ for~$s>4$. To extract $S_\ell^{(I)}(s)$ we project the amplitude as usual, by multiplying by the appropriate Legendre polynomial and integrating over the scattering angle. Obviously, this is a linear operation so that all the $S_\ell^{(I)}(s)$ can be explicitly written as linear combinations of the constants $a_{nm}$ and $b_{nm}$. 

For numerical explorations we replace $\infty$ in~\eqref{ansatz} by some large number $N_{max}$. Then we can simply explore the space of possible S-matrices by extremizing various physical observables subject to the unitarity constrains while making sure the values converge as we increase the cut-off $N_{max}$ as well as other numerical cut-offs such as the number of grid points where we check unitarity and how many spins we impose it on, see~\cite{Paulos:2017fhb} for details.~\footnote{In practice, we bound all the spins up to $\ell_{max}=20$ for the highest $N_{max}$ we use. The asymptotic large-spin behaviour is also bounded. Note that our setup here is quite distinct from other treatments -- based on Roy equations~\cite{Ananthanarayan:2000ht} for instance -- which typically include the very first few spins only.} 

So far this is totally general and valid for any unitary relativistic quantum theory with $O(3)$ vector particles. To zoom in on pions we need further input. A lot is known about pions as nicely reviewed in \cite{Tanabashi:2018oca,Pelaez:2015qba,Caprini:2005zr,Colangelo:2001df,Ananthanarayan:2000ht} and important hints can be obtained from (chiral perturbation) theory and from experiment. 

First of all we have experimental data. 
There are clear resonances in the spin $\ell=1$ and spin $\ell=2$ corresponding to the so-called $\rho$ and $f_2(1270)$ particles and a much broader resonance for spin $\ell=0$ corresponding to the $\sigma$ particle. Normally, one associates these resonances to abrupt changes in the phase shift which crosses an odd multiple of $\pi/2$.
A better characterization of these resonances is as zeros of the corresponding  partial waves. 
For example, the $\rho$ particle resonance will have a complex mass $m_\rho$ such that~\footnote{Unitarity from $4m_\pi^2$ until the first inelastic threshold must be exactly saturated. In that range $S_\ell^{(I)}(s)\hat S_\ell^{(I)}(s)=1$, where $\hat S$ stands for $S$ in the second-sheet. This functional relation can be analytically continued everywhere. As such~\eqref{position} states that a resonance corresponds to a pole in the second-sheet. A similar discussion for $1{+}1$ dimensions is in~\cite{Doroud:2018szp}.}
\begin{equation}
S_1^{(1)}(m_\rho^2)=0 \,. \label{position}
\end{equation}
Let us emphasize that a free theory has $S_\ell^{(I)}=1$ (and not zero!) so a resonance is quite a strong effect, very far from a free theory. At the same time, for pions, we do have two very important points located at sub-threshold (unphysical) values of $s$ where the weak coupling conditions (often referred as Adler's zeros~\cite{Adler:1964um})
\begin{equation}
S_0^{(0)}(s_0)=1 \qquad \text{ and } \qquad S_0^{(2)}(s_2)=1 \,, \label{weakS}
\end{equation}
	hold. The same weak coupling conditions can be imposed as zeros in the corresponding partial wave amplitudes as it follows from their definition $S_\ell^{(I)}=1+2i\sqrt{1-4/s}\,\mathcal{T}_\ell^{(I)}$. Indeed, leading 	order chiral perturbation theory~\cite{Weinberg:1966kf} predicts 
\begin{equation}
\mathcal{T}_0^{(0)}=\frac{2s-1}{32 \pi f_\pi^2},\quad \mathcal{T}_0^{(2)}=\frac{2 -s}{16 \pi\,f_\pi^2},\quad \mathcal{T}_1^{(1)}=\frac{s-4 }{96 \pi\,f_\pi^2},
\label{chitree}
\end{equation}
where $f_{\pi}$ is the pion decay constant.~\footnote{Higher loop corrections depend on the choice of constants extracted from the low energy data we want to constraint. Taking them as input would make our arguments sort of circular hence we prefer to stick to leading order here.  }
From this we read off the tree-level predictions $s_0=1/2$ and $s_2=2$. 

\begin{figure}[t]
\centering
\includegraphics[scale=0.25]{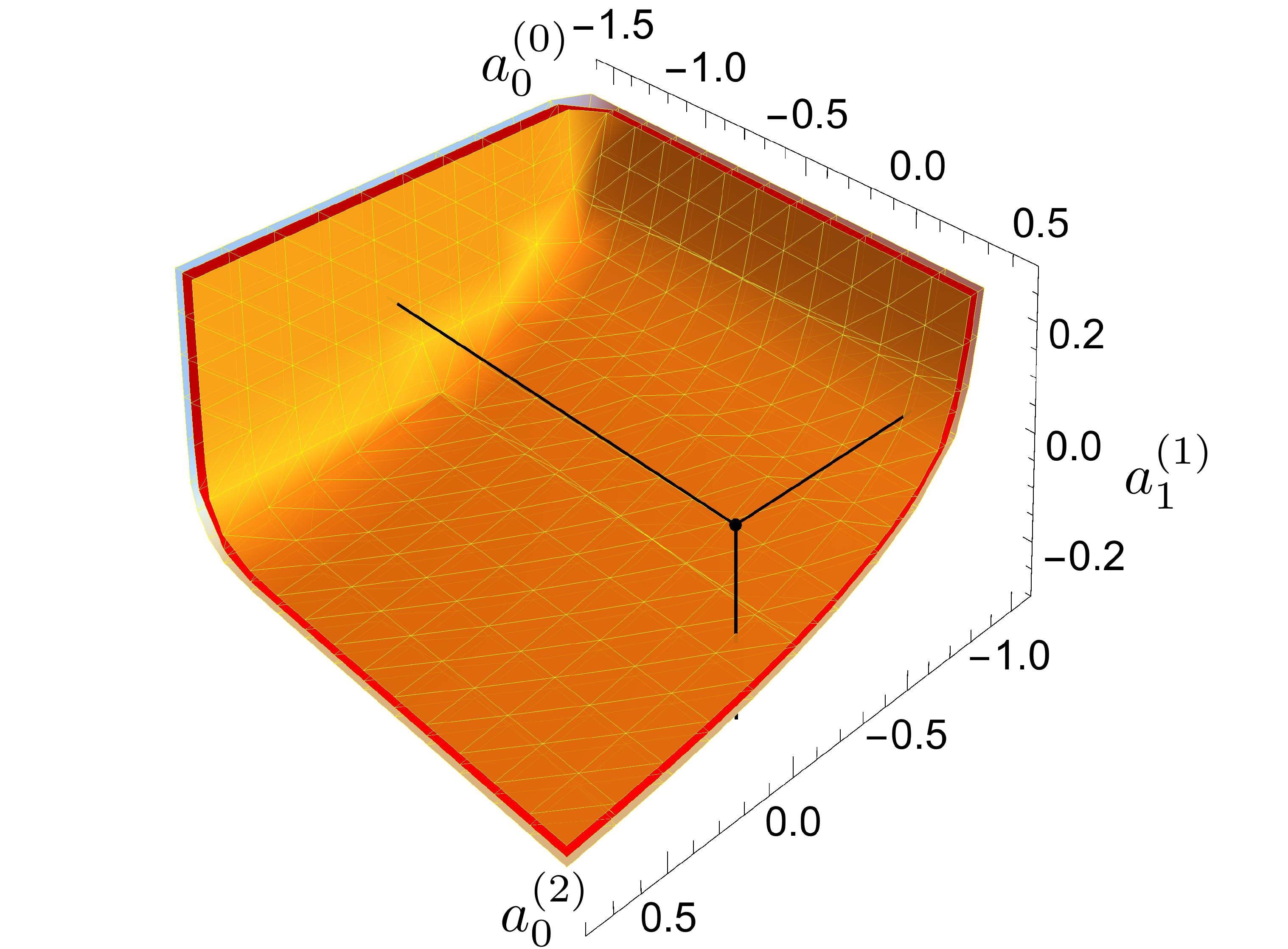}
\caption{Exploration of the minimum values scattering lengths can take. The three surfaces here correspond to~$N_{max}=12,14,16$ (orange, red, light-blue). 
The fact they are almost indistinguishable is the sign of their very good convergence.
The QCD values (Table~\ref{tableExp}) are represented by a dot here (the errors are smaller than the size of the dot) and are of course well within the allowed region where scattering lengths live.
}
\label{fig3DFirst}
\end{figure}

Finally, at low energy we have the expansion of the partial wave amplitudes close to   threshold 
\begin{equation}
\Re[\mathcal{T}_\ell^{(I)}]=k^{2\ell} \left[ a_\ell^{(I)}+b_\ell^{(I)} k^2 + \mathcal{O}(k^4) \right]\nonumber
\end{equation}
where $k=\sqrt{s/4-1}$ is the center of mass momentum,~$a_\ell^{(I)}$ are the scattering lengths and $b_\ell^{(I)}$ are the effective ranges. Since we set $m_\pi=1$, all these quantities are dimensionless: in table~\ref{tableExp} we summarize their experimental values.
\begin{table}[h]
\begin{tabular}{|l|l|l|}
\hline
I & $\mathcal{O}(k^0)$ & $\mathcal{O}(k^2)$ \\
\hline
0 & $a_0^{(0)}=0.2196 \pm 0.0034$ & $b_0^{(0)}=0.276\pm 0.006$ \\
2 & $a_0^{(2)}=-0.0444 \pm 0.0012$ & $b_0^{(2)}=-0.0803\pm 0.0012$\\
1 & & $a_1^{(1)}=0.038 \pm 0.002$\\\hline
\end{tabular}
\caption{Experimental determination of scattering lengths and effective ranges of $\pi$-$\pi$ scattering up to spin one~\cite{Nagels:1979xh,Batley:2010zza}.}
\label{tableExp}
\end{table}




\section{Numerical explorations}
We first ask what is the allowed region in the three-dimensional space of scattering lengths $\{a_0^{(0)},a_1^{(1)},a_0^{(2)} \}$.
It turns out they are all bounded from below~\cite{Wigner:1955zz}.~\footnote{ The boundedness of scattering lengths comes from their relation to time delays at low energies; while positive and arbitrarily large time delays are clearly allowed (and do show up in our numerics as resonances arbitrarily close to threshold trapping the particles), negative large time delays could lead to causality violations if no bound states are present and hence the existence of these lower bounds. Similar (but not identical) bounds were previously explored in~\cite{Lopez:1975gk,Lopez:1975wf,Lopez:1974cq,Caprini:1980un}.}
The boundary of the allowed region is shown in Fig.~\ref{fig3DFirst}. It has  a smooth tip, a point of highest curvature; we do not know  if it corresponds to any physical theory. The black dot represents the QCD experimental values from table~\ref{tableExp}. We see that it is well inside the allowed region. Indeed, we can study the  various phase shifts of the extremal solutions  along the boundary and realize these do not resemble QCD. This is not surprising because we are not imposing anything yet about the chiral symmetry breaking physics which pions describe.


\subsection*{The lake}

To do so, we would like to  impose the existence of the two chiral zeros described in~\eqref{weakS}. However, these zeros appear in the unphysical region $s<4$ and their position cannot be measured experimentally. So our next investigation aims at getting some hints about their position. 

\begin{figure}[t]
\centering
        \includegraphics[scale=0.14]{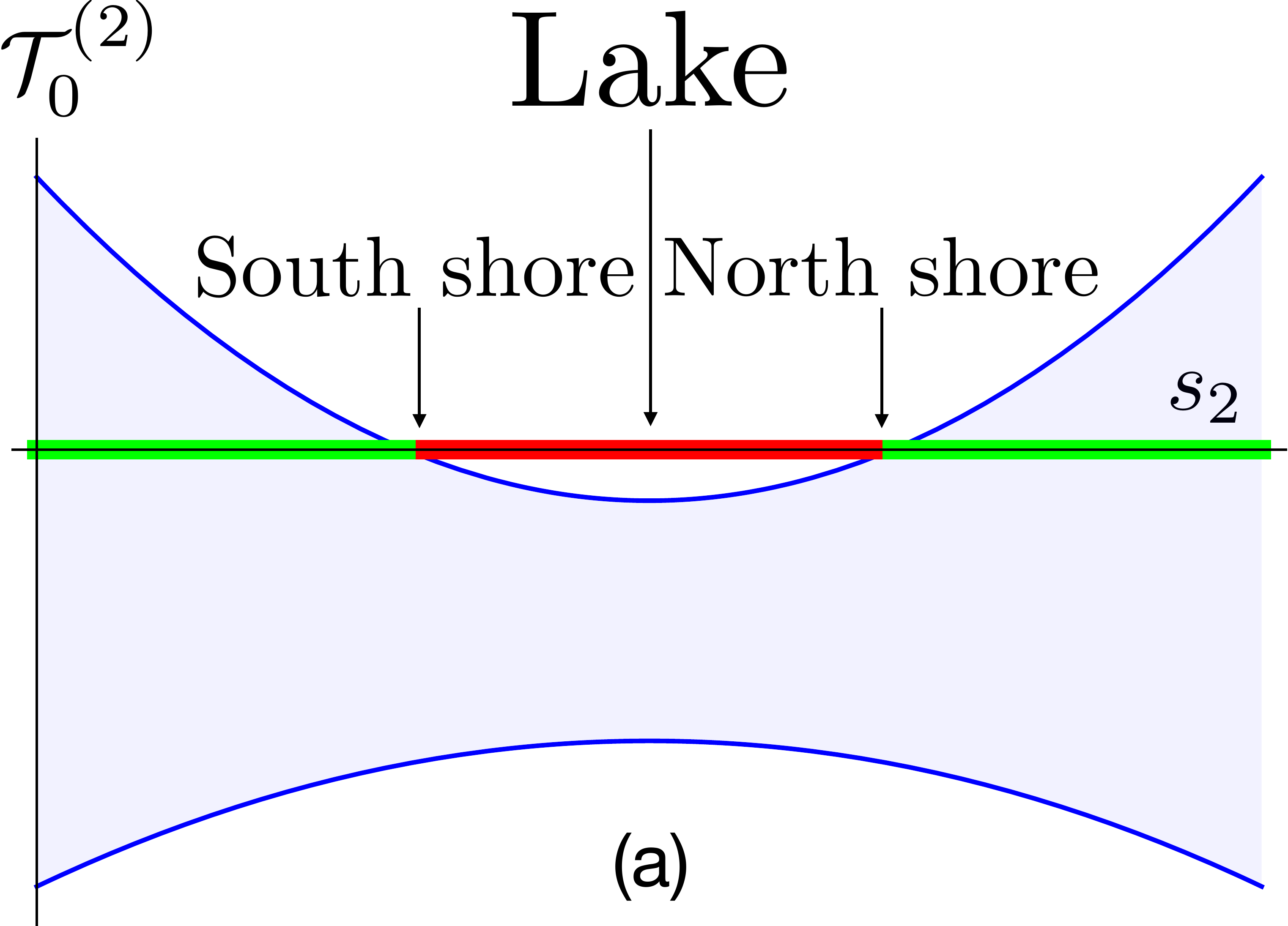}
        \includegraphics[scale=0.14]{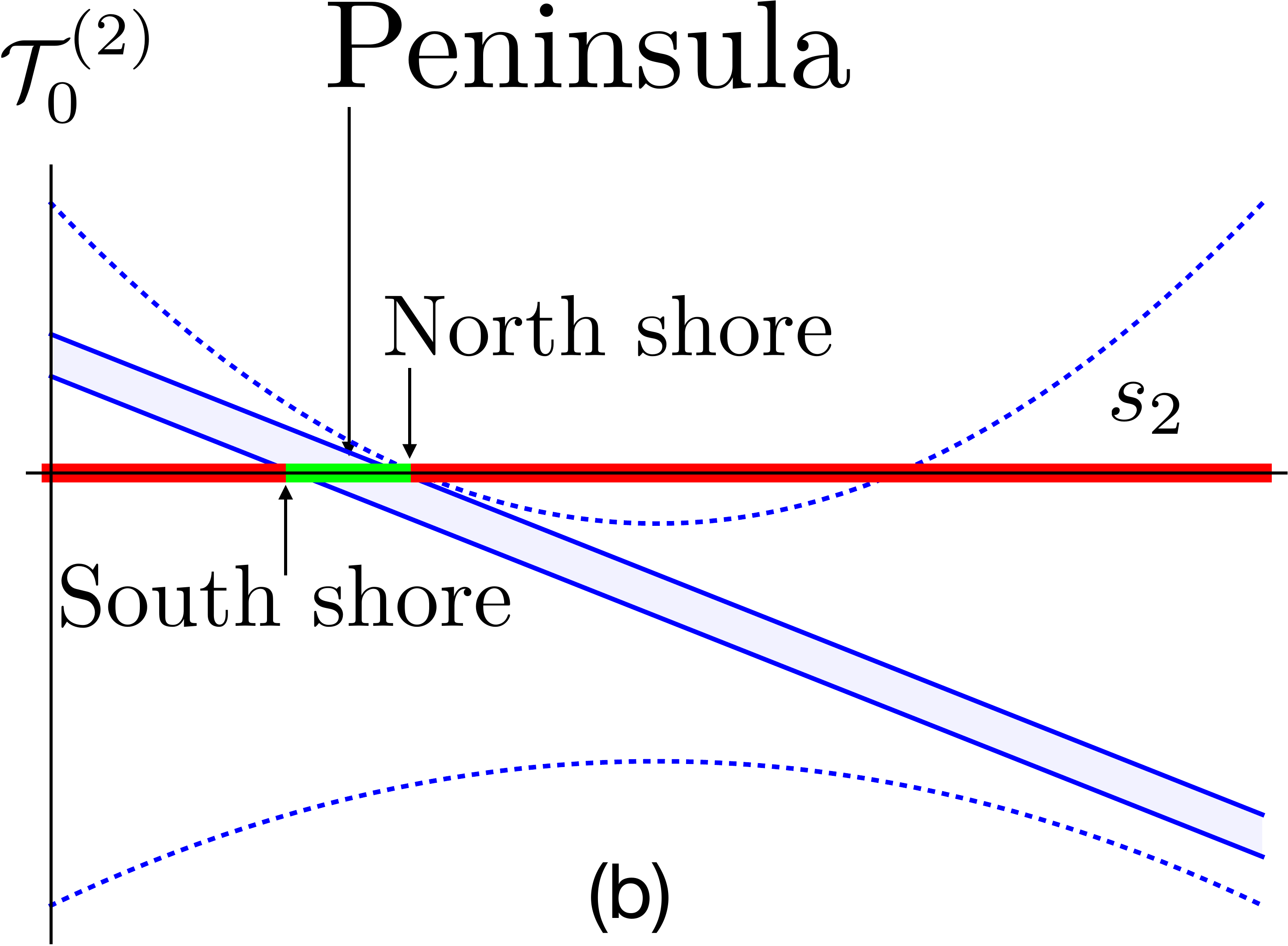}
 \caption{Schematic picture of the lake, panel (a), and of the peninsula determination, (b).  In (a) the solid blue lines enclose the allowed region for the amplitude $\mathcal{T}_{0}^{(2)}$ when we impose the chiral zero condition at $s_0=1/2$ and the $\rho$ resonance. The same region is shown in panel (b) by dashed blue lines. The solid blue lines in (b) embraces the allowed region when the three scattering lengths are set to the experimental values within errors. 
 In both panels we denote in red the region where we cannot fix $s_2$ and in green where we can.}
\label{sections}
\end{figure}

\begin{figure}[t]
\centering
\includegraphics[scale=0.30]{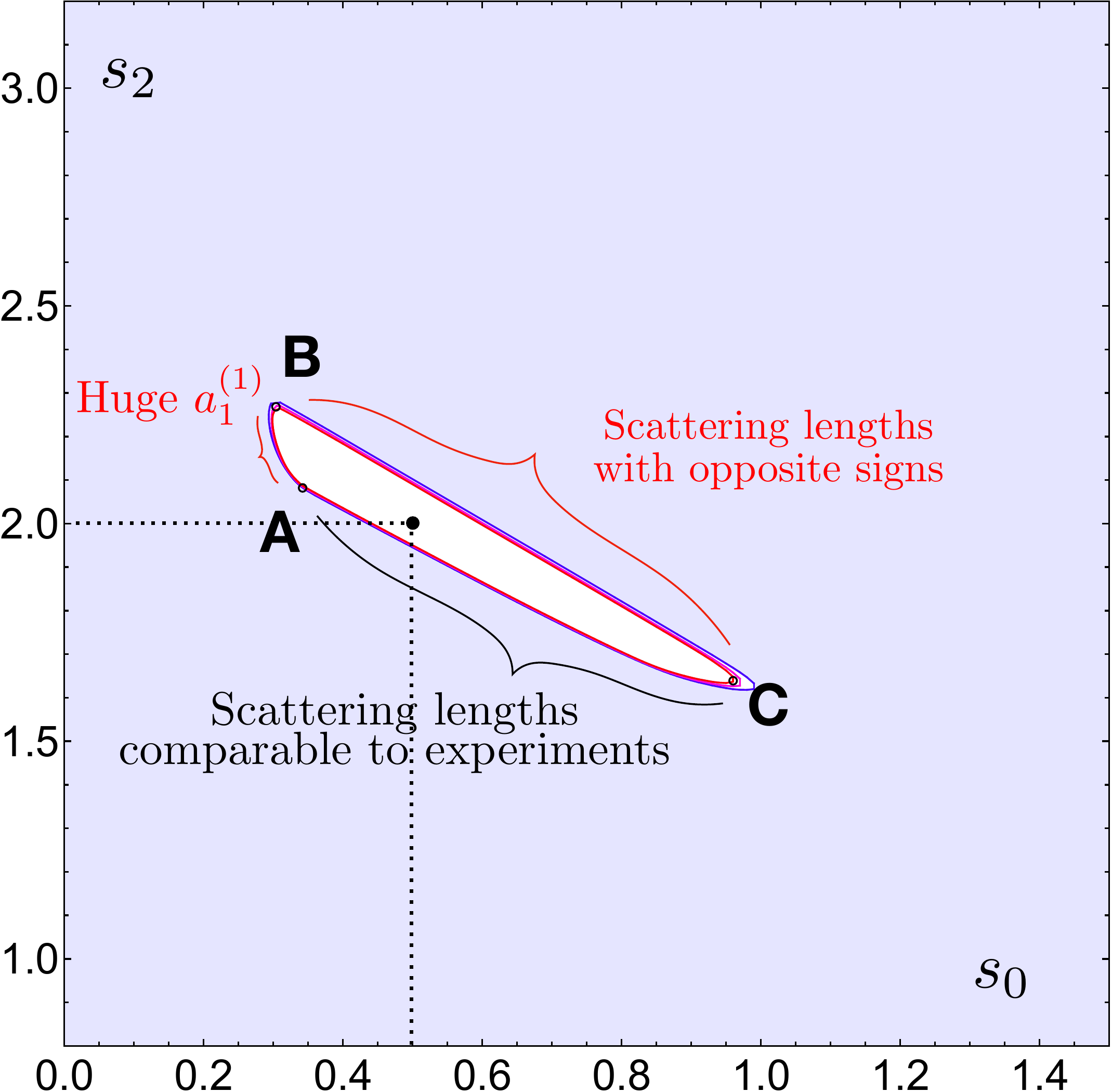}
\caption{Pion Lake: the white region is the exclusion area in the plane $(s_0, s_2)$ when we fix the $\rho$ resonance. The black point corresponds to tree-level chiral perturbation theory~\eqref{chitree} which is now excluded. We show the shape of the lake for three different $N_{max}=12,13,14$ (blue to red): convergence is guaranteed by the three curves almost overlapping. }
\label{figLake}
\end{figure}

We start by fixing one chiral zero $s_0$ in the singlet channel amplitude $\mathcal{T}_0^{(0)}(s)$ and the vector $\rho$ resonance with mass parameters $m_\rho \simeq (5.5+0.5\,i) $.~\footnote{The condition $\mathcal{T}_0^{(0)}(s_0)$=0 sets one linear constraint on the constants $a_{nm}$ and $b_{nm}$ in the ansatz~\eqref{ansatz} which allows us to solve for one of these constants. The resonance condition eliminates two more constants because the real and imaginary part of eq.~\eqref{position} are different constraints.} Then we maximize and minimize the symmetric channel amplitude $\mathcal{T}_0^{(2)}(s)$ for $0 <s  <4$ obtaining the blue solid curves in Fig.~\ref{sections}(a) (here represented $s_0=1/2$). We learn that there are regions where we cannot impose a chiral zero $s_2$ since the maximum and minimum are both negative there (red segment). Therefore,  we can simultaneously impose the weak coupling conditions $\mathcal{T}_0^{(2)}(s_2)=0$ and $\mathcal{T}_0^{(0)}(s_0)=0$ only when the upper boundary is positive (green segment). Repeating this game for various $s_0$ allows us to exclude a full region in the $(s_0,s_2)$ plane. This region, depicted in Fig.~\ref{figLake} is what we dub the~\textit{pion Lake}.

Notice that, while the amplitude maximizing $\mathcal{T}_0^{(2)}(s)$ is unique, 
there can be many amplitudes having the same zero $s_2$. 
On the other hand, the intersection of the upper boundary of the allowed region with the $s_2$ axis selects a unique amplitude since a zero there coincides with the maximum possible value at that point. This uniqueness of the boundary theories is the same sort of uniqueness found and thoroughly explored in the conformal bootstrap~\cite{ElShowk:2012hu}. 

Remarkably, the prediction $(s_0,s_2)=(1/2,2)$ of leading order chiral perturbation theory~\eqref{chitree} is clearly excluded. What the lake is telling us is that we need to be considerably far from that weak coupling point to be able to include such a strong coupling phenomena as a resonance. 
Therefore, we can  think that the boundary of the lake corresponds to theories that are \textit{as free as possible} given that they contain the $\rho$ resonance. 

If we were to take the complex $\rho$ mass to be larger and larger (i.e. further away in the Mandelstam plane) the lake (i.e. the excluded region) would become thinner and thinner eventually becoming a line segment passing through the chiral perturbation theory point and very well fitted by our numerics to $s_2-2+4/5(s_0-1/2)=0$. Curiously this line of zeros would correspond to a tree level amplitude $A(s|t,u)\propto s - m_\pi^2 \alpha$; $\alpha=1$ yields~\eqref{chitree}. It would be interesting to further explore this line segment analytically; it should be related to an interesting line of perturbative field theories. 

\begin{figure}[t]
\centering
\includegraphics[scale=0.241]{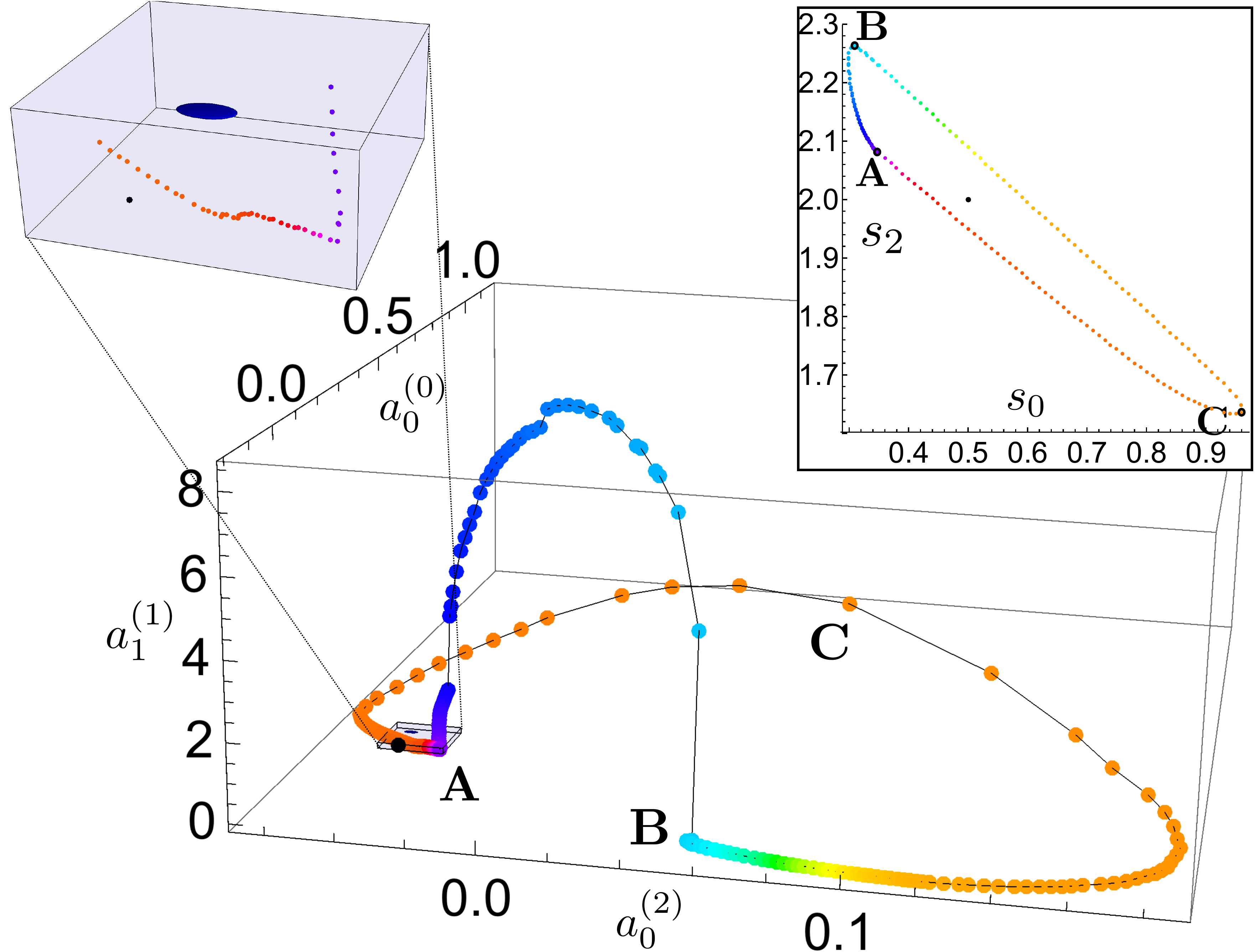}
\caption{Scattering lengths orbit around the lake. The colors in the inset and in the panel match in order to help following the orbit as we move around the lake. 
In the left inset we zoom in on the $A$-kink: the black dot corresponds to the tree-level chiral theory and the ellipsoid 
to the experimental values for QCD. All the curves shown are obtained at fixed $N_{max}=14$.
The small non-smoothness of the $A$-$B$ arc is a  numerical artefact (in any case it occurs in a region where $a_1^{(1)}$ is huge, far from the pion physics we are interested in).}
\label{fig3DAFancy}
\end{figure}

We next ask whether anything interesting happens along the lake shore for these freest possible theories. 
In Fig.~\ref{fig3DAFancy} we follow the scattering lengths around the lake. The points $A$ and $B$ -- also in Fig.~\ref{figLake}, correspond to cusps both in the the scattering length orbit and in the lake shape, as opposed to $C$ which is a smooth transition point. It is remarkable that along the shore $A$-$C$ and, in particular, at the kink $A$ the sign of the scattering lengths match the experimental ones being also close in magnitude to them (see the left-inset in Fig.~\ref{fig3DAFancy}).

\subsection*{The peninsula}

\begin{figure}[t]
\centering
\includegraphics[scale=0.22]{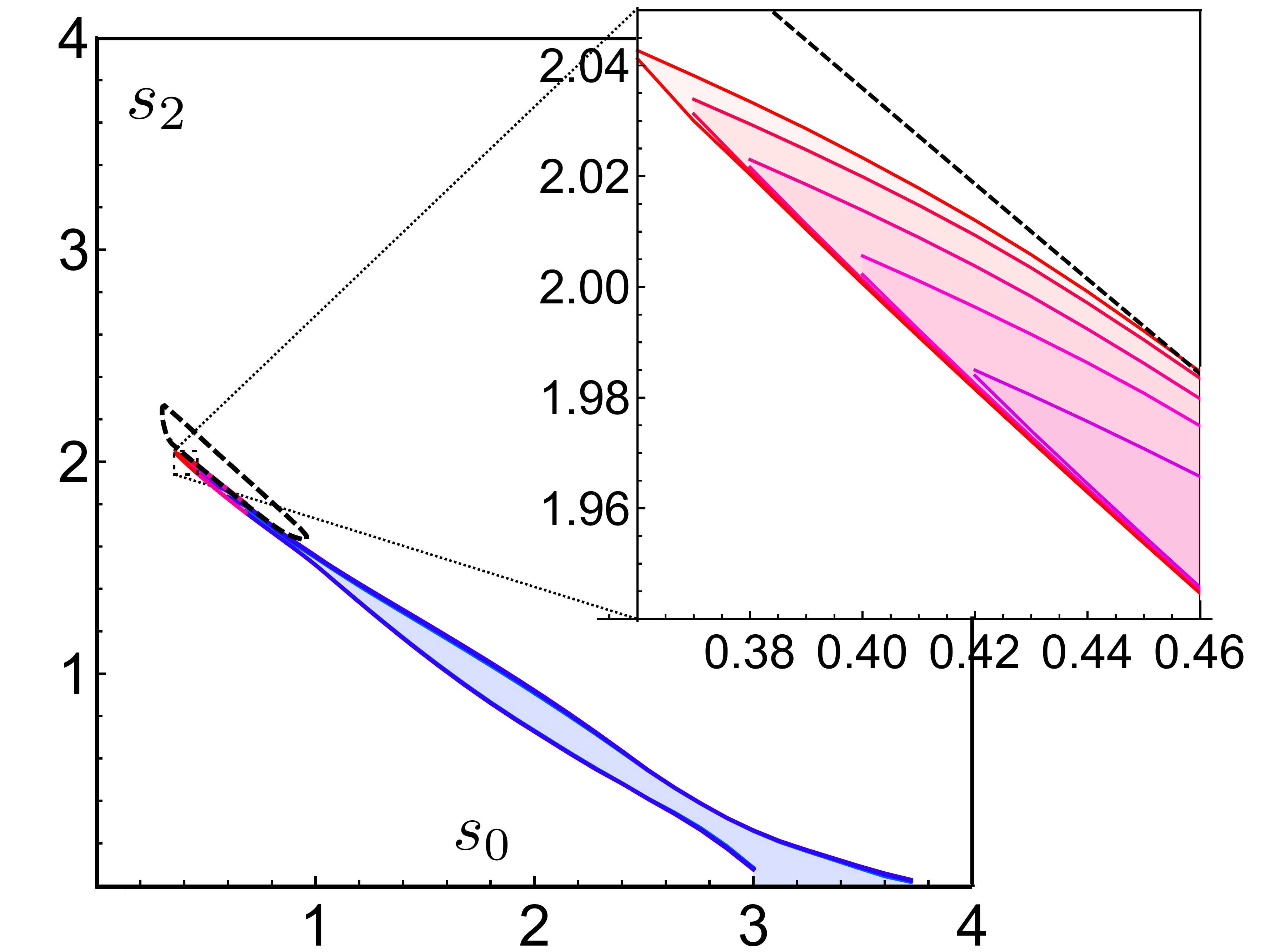}
\caption{\textit{Pion peninsula}: we fix the $\rho$ resonance and the inequalities given by the experimental intervals for the three scattering lengths. The colored region is the one allowed by unitarity and different colors correspond to different $N_{max}$ from 12 to 20 as we go from \textit{blue} to \textit{red}. The dashed contour encloses the \textit{pion lake}. }
\label{PeninsulaFigure}
\end{figure}

Motivated by these explorations, we consider a third exploratory game where we now impose the $\rho$ resonance condition plus the scattering length inequalities $|a_0^{(0)}{-}0.2196|< 0.034$ etc given in Table~\ref{tableExp}. These upper and lower bounds for each scattering length are simply 6 additional constraints to add to the many unitarity conditions we have already. We repeat the procedure above of maximizing and minimizing the value of the amplitude in the symmetric channel $\mathcal{T}_0^{(2)}(s)$ with a zero $s_0$ imposed in the singlet channel $\mathcal{T}_0^{(0)}(s)$. 
The schematic representation of the result is in~Fig.~\ref{sections}(b).
We note that while before we excluded a small segment, now the maximum and minimum conditions exclude \textit{all but} a tiny region for possible positions $s_2$ of the second chiral zero since only in a very small segment is the minimum negative and the maximum positive! Scanning over various $s_0$ we thus construct a full region in the $(s_0,s_2)$ plane which is now mostly excluded. It is represented in Fig.~\ref{PeninsulaFigure} and dubbed as the \textit{pion peninsula} for obvious reasons.

It would be very interesting to see how the excluded region (the exterior of the peninsula) shrinks to the lake in the limit when the inequalities imposed on the scattering lengths become looser.

\begin{figure}[t]
\centering
\includegraphics[scale=0.23]{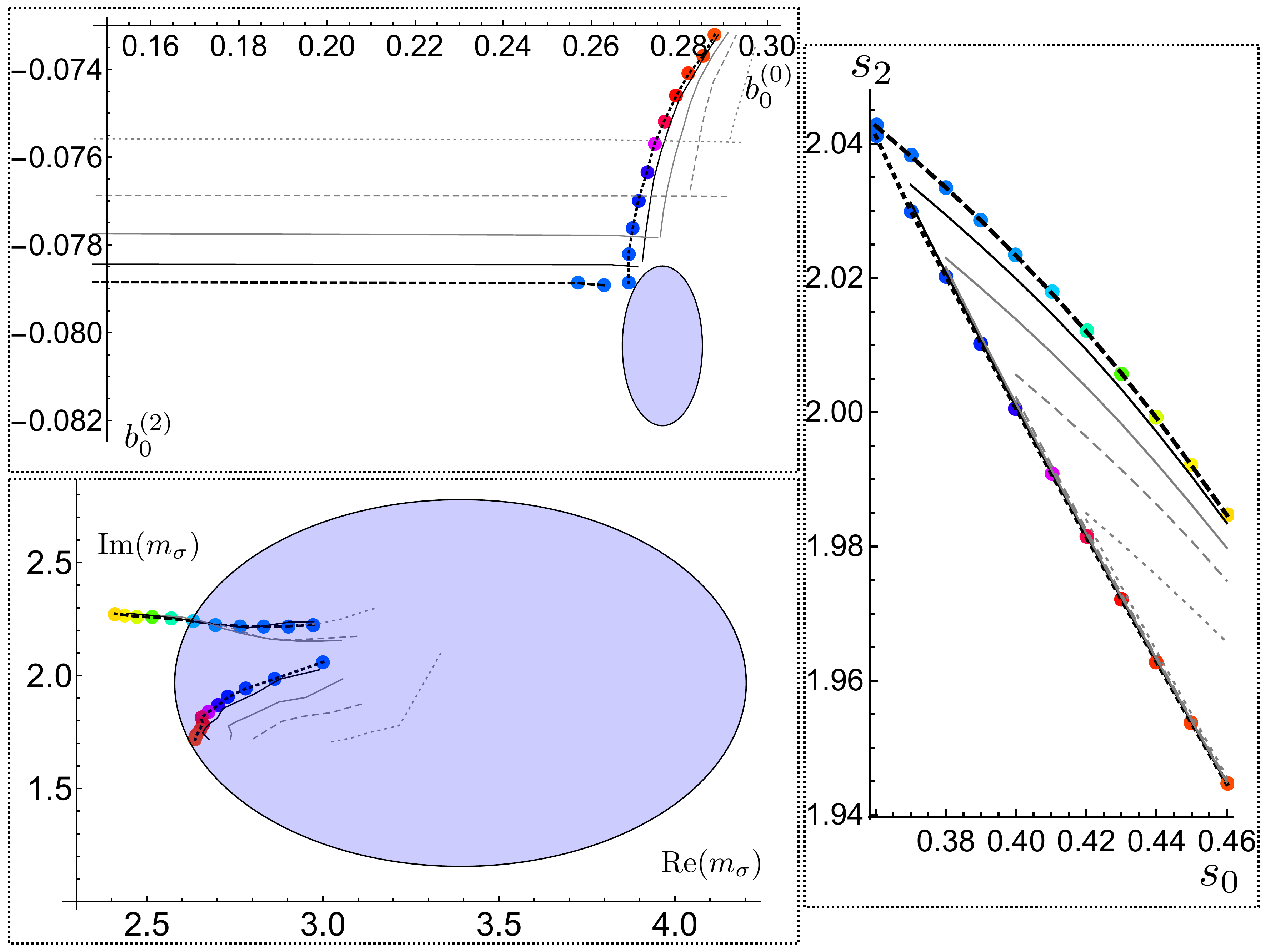}
\caption{Trajectory of the effective ranges in the $(b_0^{(0)}, b_0^{(2)})$ plane (upper left panel), and of the spin-0 zero in the $(\text{Re}(m_\sigma),\text{Im}(m_\sigma))$ plane (lower left panel) as we travel on the boundary of the peninsula (right panel), at fixed $N_{max}=20$. The blue elliptic regions represent the experimental values and their uncertainty. In gray from lighter to darker the same trajectories for $N_{max}=16,...,19$.}
\label{figbs}
\end{figure}

In Fig.~\ref{figbs} we plot the effective ranges  $b_0^{(0)}$, $b_0^{(2)}$ as we move around the peninsula boundary.
Very nicely, we note that at the tip of the peninsula the orbit passes close to their experimental values in Table~\ref{tableExp}. 
Besides, we observe the emergence of a first zero in $S_0^{(0)}$ that we identify as the $\sigma$ resonance. 
It too approaches its experimental position as we go to the tip of the peninsula.

\subsection*{The kink}
By now, we have a few natural candidates for reasonable chiral zeroes' positions. We could take the values along the shore of the pion lake which are closest to experiment; or the sharp kink we observe along that shore; or we could take the tip of the pion peninsula. 
In this concluding section we fix the chiral zeros to the tip of the peninsula at our largest $N_{max}=20$ value as illustrated in Fig.~\ref{PeninsulaFigure}. 

With these  chiral zeros fixed, we repeated the analysis of the possible allowed values for the three scattering lengths, as in Fig.~\ref{fig3DFirst}. We observed that with these extra constraints the resulting space is much more interesting and intricate. 
We refrain from depicting it here since we first want to make sure that our numerics have properly converged close to all the various kinks and edges which comprise its very rich boundary.
 
 \begin{figure}[t]
\centering
\includegraphics[scale=0.39]{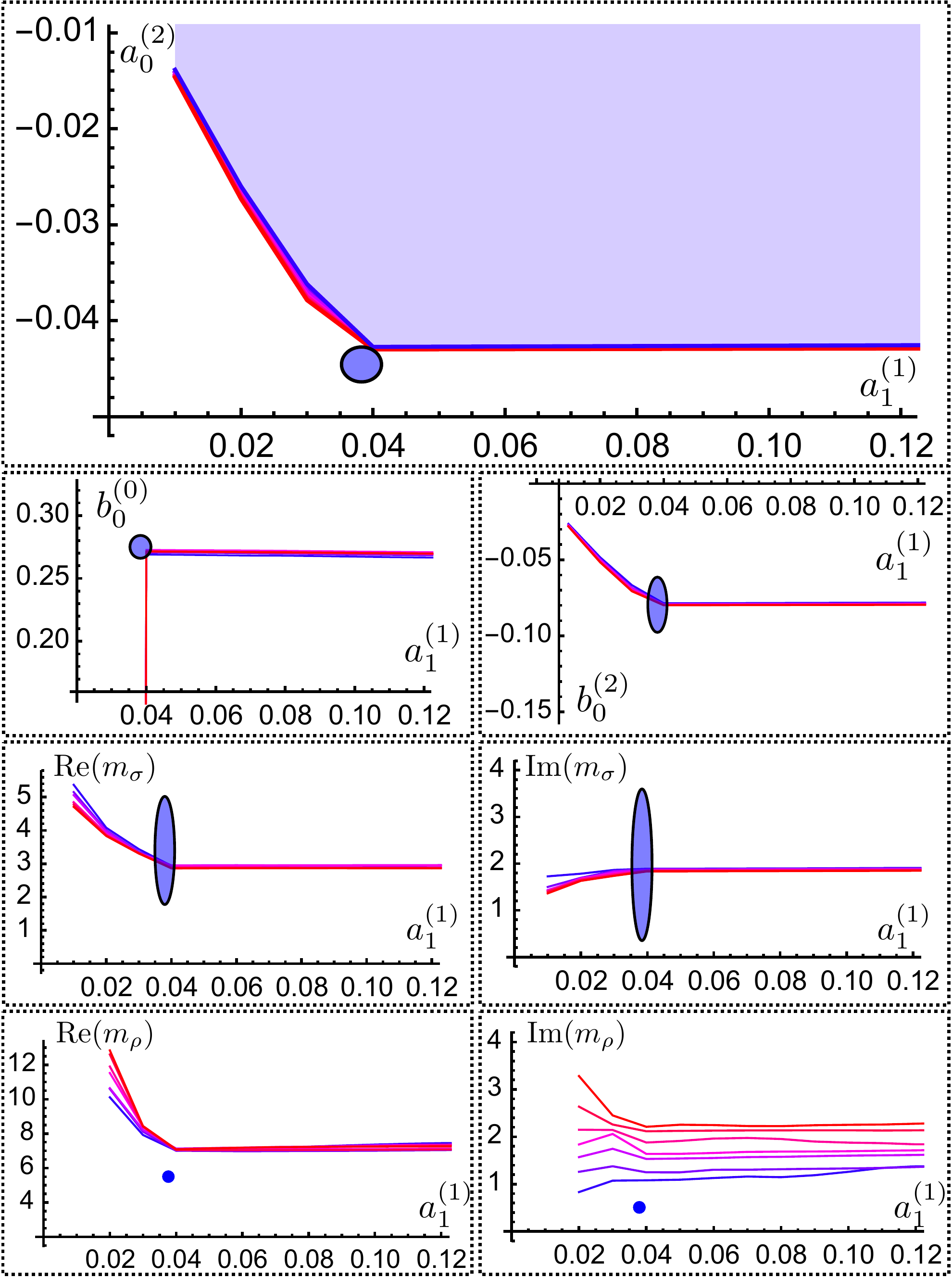}
\caption{Top-right panel: allowed region in the $(a_0^{(2)}, a_1^{(1)})$ plane at fixed $s_0=0.36$, $s_2=2.04$ and $a_0^{(0)}=0.2196$. All the other panels show the effective ranges and the resonance masses 
along the same boundary as a function of $a_1^{(1)}$. Their experimental values lie in the blue ellipses. $N_{max}$ goes from $14$ to $20$ (blue to red).}
\label{figkink}
\end{figure}

There seem to be, for example, two cusps in this three dimensional space (one corresponding to the free theory and another with scattering lengths roughly twice as large as those measured in nature). The two cusps are connected by a sharp edge where the QCD scattering lengths seems to lie. To illustrate this, we consider in Fig.~\ref{figkink} a two-dimensional section of the allowed space intersecting this sharp edge at fixed  $a_0^{(0)}=0.2196$, the expected experimental value for this scattering length. As illustrated in this figure, not only do we have a very sharp kink but its location is extremely close to the experimental values. 


Moreover, if we follow other quantities as the effective ranges $b_0^{(0)}$, $b_0^{(2)}$ or the complex mass $m_\sigma$ of the emerging spin-zero resonance as a function of $a_1^{(1)}$, they all show a kink compatible with their experimental values! We even observe the emergence of a zero in $S_1^{(1)}$ which we would like to identify with the $\rho$ resonance. (As seen in the figure, its real part is actually quite close to the experimental one but its imaginary part is off.)~\footnote{In our setup, the boundary of our S-matrix theory space will typically saturate unitarity. However, at some high energy we have sizeable inelasticities in pion scattering most notably at the $K\bar K$ threshold. As an instructive exploration we imposed -- by hand -- reasonable inelasticity in the unitarity constraints starting at this threshold and observed that most of the low energy parameters such as the ones plotted here are stable and almost do not change. (Similar games were played in two dimensions in \cite{Cordova:2018uop} with similar conclusions.) To rigorously take these physical processes into account we would have to scatter $K$-ons as well and consider multiple pion scattering processes.}

We find it remarkable that the theories lying at the boundary of the allowed space have any resemblance to those in the real world, with resonances emerging dynamically and with various physical parameters extremely close to the values observed in nature. Of course, a more quantitative comparison (for the $\rho$ particle life time for instance) would entail a more detailed study of the convergence of all phase shifts, exploration of various chiral zeroes scenarios, etc.\footnote{Notice that  the pion S-matrix is far from unique. 
For starters, our assumptions are common to all QCD-like theories with fixed symmetry breaking pattern in the IR, but different gauge group. Moreover, for each gauge group,
 there is  a continuous family of theories as a function of $\Lambda_{QCD}/m_\pi$. The situation is not dissimilar from the lattice formulation where one needs to fix the gauge group and set the bare masses of the quarks.
 However, it is still hard for  lattice simulations to work with physical pion masses~\cite{Briceno:2017qmb}. It would be interesting to use our framework to study the scattering of particles in a regime accessible to the current lattice simulations.} We hope to report of these explorations soon.


%

\section{Conclusions}

We started exploring the space of low energy parameters of massive QFT in $3+1$ dimensions that have a low energy effective field theory description, focusing in particular on QCD. Inspired by the plethora of experimental and theoretical results, we find that imposing additional conditions as the presence of a resonance or inequalities on the scattering lengths the space of low energy parameters shows a very non-trivial structure with lakes, peninsulas and kinks. For the reader's convenience, here is a telegraphic summary table:

\begin{table}[h]
\begin{tabular}{|l|l|}
\hline
\textbf{input} & \textbf{output} \\
\hline
nothing imposed & absolute bound of scatt. lengths Fig.~\ref{fig3DFirst}  \\
\hline
$\rho$ resonance & \textit{pion lake} exclusion for chiral zeros Fig.~\ref{figLake}\\
 & scatt. lengths around the lake in Fig.~\ref{fig3DAFancy} \\
 \hline
$\rho$ + exp. & \textit{pion peninsula} Fig.~\ref{PeninsulaFigure}\\
  scatt. lengths & new spin 0 resonance + eff. ranges Fig.~\ref{figbs}\\
\hline
$s_0$, $s_2$ & Intricate allowed $a_\ell^{(I)}$ 3D space as briefly\\
&  described in the text~\cite{toappear}\\
\hline
$s_0$, $s_2$, $a_0^{(0)}$& kink in $(a_0^{(2)}$, $a_1^{(1)})$ Fig.~\ref{figkink} \\
& kinks in $b_0^{(0)}$, $b_0^{(2)}$, $m_\sigma$ as function of $a_1^{(1)}$\\
\hline
\end{tabular}
\label{summary}
\end{table}
 
It would be very interesting to extend this study to the case of massless pions, \emph{i.e.} exact Goldstone bosons and to other interesting setups with some symmetry breaking pattern, both in four or lower dimensions. (In two dimensions, the $O(N)$ bootstrap was recently addressed in \cite{He:2018uxa,Cordova:2018uop,Paulos:2018fym} where contact with known integrable models was made.)
We hope to report on progress in these directions soon.

One of the main outcomes of the discussion contained in this Letter is the constraint on the position of the chiral zeros in $\pi$-$\pi$ scattering, obtained in a consistent way using our setup -- see Fig.~\ref{PeninsulaFigure}. An obvious next question which we are currently exploring is whether imposing the chiral zeros in the allowed region is enough to select a theory with phase shifts resembling the experimental ones and from which we can extract a resonance spectrum compatible with the physical one. 



\begin{acknowledgments}
We would like to thank N. Arkani-Hamed, C. Bercini, M. Bianchi, L. Cordova, S. Frautschi, D. Gaiotto, A. Homrich, S. Komatsu, M. Kruczenski, M. Paulos, B. van Rees and K. Zarembo for discussions. We wish to thank J. Toledo for collaboration in the early stage of this work. 
We are indebted to J.R. Pelaez for sharing with us the experimental phase shifts for $\pi$-$\pi$  scattering.  
We thank J.R. Pelaez and A. Pilloni for useful comments on the draft.
Research at the Perimeter Institute is supported in part by the Government of Canada through NSERC and by the Province of Ontario through MRI. This research received funding from the Simons Foundation grants 
JP:\#488649 and  PV:\#488661
(Simons collaboration on the Non-perturbative bootstrap) and FAPESP grant 2016/01343-7 and 2017/03303-1. JP is supported by the National Centre of Competence in Research SwissMAP funded by the Swiss National Science Foundation.
\end{acknowledgments}

\appendix

\nocite{*}

\bibliography{thepionlakev1}

\end{document}